# First-Principles Studies of Stacking Fault Energies in Ternary Magnesium Alloys

Qiwen Qiu, Stephen Yue, Jun Song[1]

*Department of Mining and Materials Engineering,*

*McGill University, Montréal, Québec, H3A 0C5, Canada*

## Abstract

Magnesium (Mg) alloys have emerged as promising materials due to their low density and high strength-to-weight ratio, offering a wide range of applications across multiple industries. Nevertheless, the inherent brittleness of Mg alloys poses a significant hurdle, necessitating innovative approaches to enhance their mechanical performance. Among the various strategies, manipulating stacking fault energy (SFE) has been a key focus, although primarily within the realm of binary alloys. This study investigates SFE in Mg alloys, focusing on ternary compositions. Utilizing first-principles DFT calculations, we analyze solute interactions and their influence on SFE, particularly in Mg-Al-X and Mg-Zn-X configurations. Predictive models are developed for estimating SFE effects, revealing solute pairs that mimic rare earth elements and show potential for improved ductility. The findings contribute to fundamental insights into Mg alloy behavior, offering practical directions for designing advanced materials with superior mechanical properties.

[1] The author to whom correspondence is to be addressed. Email: jun.song2@mcgill.ca; Tel: +1 514-436-7649

## 1. Introduction

In the realm of materials science and engineering, the allure of Mg alloys has grown significantly in recent years, owing to their remarkable combination of low density, excellent strength-to-weight ratio, and good biocompatibility [1-3]. The pursuit of high-performance Mg alloys is underscored by their potential to revolutionize the design and manufacturing of lightweight structural materials, offering a compelling solution for achieving enhanced fuel efficiency and reduced environmental impact [4, 5]. Despite their unique advantages, Mg alloys exhibit inherent brittleness, posing a substantial challenge to their widespread adoption [6]. This brittleness is attributed to several factors, among which the limited independent slip systems and high anisotropy of hexagonal close-packed (hcp) structure play a crucial role [7, 8]. In a hcp structure, there exist solely two independent slip systems within the basal plane, while other deformation modes, including twinning and non-basal slips, suffer from high activation barriers which render them difficult to operate at room temperature [9-12]. This results in the available deformation modes effectively falling short of the requisite five independent modes for achieving compatible deformation, thus poor formability at room temperature [10, 11]. Additionally, it has been demonstrated that the non-basal pyramidal I and II $< c + a >$ dislocations may undergo transformation into basal dislocation structures, which would further contribute to the brittleness of Mg alloys [13, 14].

Numerous efforts have been carried out, attempting to overcome the afore-mentioned challenges, in the design of Mg alloys. Various methods such as grain refinement and solid solution alloying have shown promise in improving the plasticity of Mg alloys. For instance, the addition of rare earth (RE) elements to Mg alloys has been found to significantly enhance elongation [15-19]. However, the exact mechanisms by which RE elements improve plasticity are still under debate. Some studies suggest that RE elements alter the $I_1$ stacking fault energy (SFE) then enables nucleation and activity of $< c + a >$ dislocations [15], while other research proposes that RE elements change the $< c + a >$ dislocation cross-slip rates, thereby influencing the plasticity [13]. Despite the difference, these studies converge in recognizing the crucial relevance of SFE in Mg alloys. In this regard, considerable research efforts have been dedicated to understanding the role of SFE, and how one may modulate SFE through alloying [9, 11, 16-20]. For example, Muzyk's investigation demonstrated that Pd and Sn have the highest reduction on SFE of Mg, which decrease the energy barrier for partial dislocations and stacking fault formation greatly [16]. Datta et al. proposed that the addition of Zn would improve the ductility of the alloy with a reduced SFE [17]. Sandlobes et al. explored the correlation between ductility and SFEs in Mg and Mg-Y alloys [11]. In Yin's study, they provided the effects of Y, Al and Zn solutes on all the pertinent SFEs in Mg and suggested that the enhanced ductility is related to the differing effects of solutes on the SFEs and dislocations of the two pyramidal planes [9]. The study conducted by Shang et al. on binary $Mg_{95}X$ revealed that the basal unstable $I_2$ SFE demonstrates an approximately inverse linear association with the equilibrium solute volume [18]. The above studies are just a few among many that contributed great merits, particularly in the front of understanding the effect of solute elements on SFE in Mg alloys. However, those previous studies on SFE focused on the effect of single solute, while in commercially significant Mg alloys, two or more solutes are often considered in the alloy design. It has been demonstrated that in ternary or quaternary alloys, the overall effect of solutes is not a simple superposition of the individual contribution from each solute [19-21]. For instance, the basal slip's unstable SFE in Mg exhibited a more pronounced decrease in systems where both Y and Zn were simultaneously introduced compared to those resulted with individual additions of either Zn or Y [22]. Additionally, transmission electron microscope images revealed an augmented separation distance between two Shockley partial dislocations in Mg-Zn-Y alloys, affirming that the concurrent incorporation of Zn and Y results in a substantial lowering of SFE [23]. Besides, solute pairs were also found to exist in Mg alloys [19]. Therefore, it is important to study the interaction between solutes, and the role of solute interaction in affecting SFE of Mg alloys.

The present study targets in addressing such need. Focusing on ternary Mg alloys, we calculated various solute-solute interaction energies and the resulted SFE in the presence of solutes. Particular attention is paid to the prediction of solute pair effects under different configurations on SFE. Based on the calculation results, several alternative elements to RE elements were come up with. We also discussed the implications of SFE change on non-basal slip activation mechanisms. The results provide critical mechanistic information to the SFE in Mg alloys and practical insights for the design of new Mg alloys with improved mechanical properties.

## 2. Methodology

First-principles density functional theory (DFT) calculations were performed to study the SFEs of Mg alloys. For these calculations, a 96-atom supercell with 12 layers and a vacuum gap of 15 Å between periodically repeated slabs was used. In this study, 16 kinds of commonly used alloying elements, including Al, Ca, Dy, Er, Gd, La, Li, Mn, Nd, Sc, Si, Sm, Ti, Y, Zstn, and Zr were considered. The generalized stacking fault energy curves (GSFEs) were calculated using the slab shearing method [15, 16] where atoms above the slip plane are displaced by a shift along Burgers vector ($\boldsymbol{b}$) relative to the atoms below, followed by relaxation in the direction normal to the slip plane. The unstable stacking fault and stable stacking fault energies, denoted as $E_{usf}$ and $E_{sf}$ respectively, can then be extracted from the GSFE curves obtained. To study the interaction behavior between solutes, another 96-atom supercell without vacuum gap was also built. For brevity, the supercells used for our calculations are illustrated in Fig. 11 and Fig. 12 in the Supplementary Information. For the study of solute interactions, the solutes were introduced by substituting Mg atoms on or at the close vicinity of the slip plane, and we can categorize these interactions according to the relative positions of the atoms within the solutes, specifically the three relative positional relationships, including two first near neighbouring positions (1NN and 1NN') and one second near neighbouring position (2NN), as illustrated in Fig. 5.1a. The solute-solute interaction and solute-stacking fault interaction approach to zero as the distance between them increase [19, 23]. That implies only the solutes near each other and also near the stacking fault may have a significant effect on SFE. Therefore, solute interactions with the separation distance beyond 2NN are not considered.

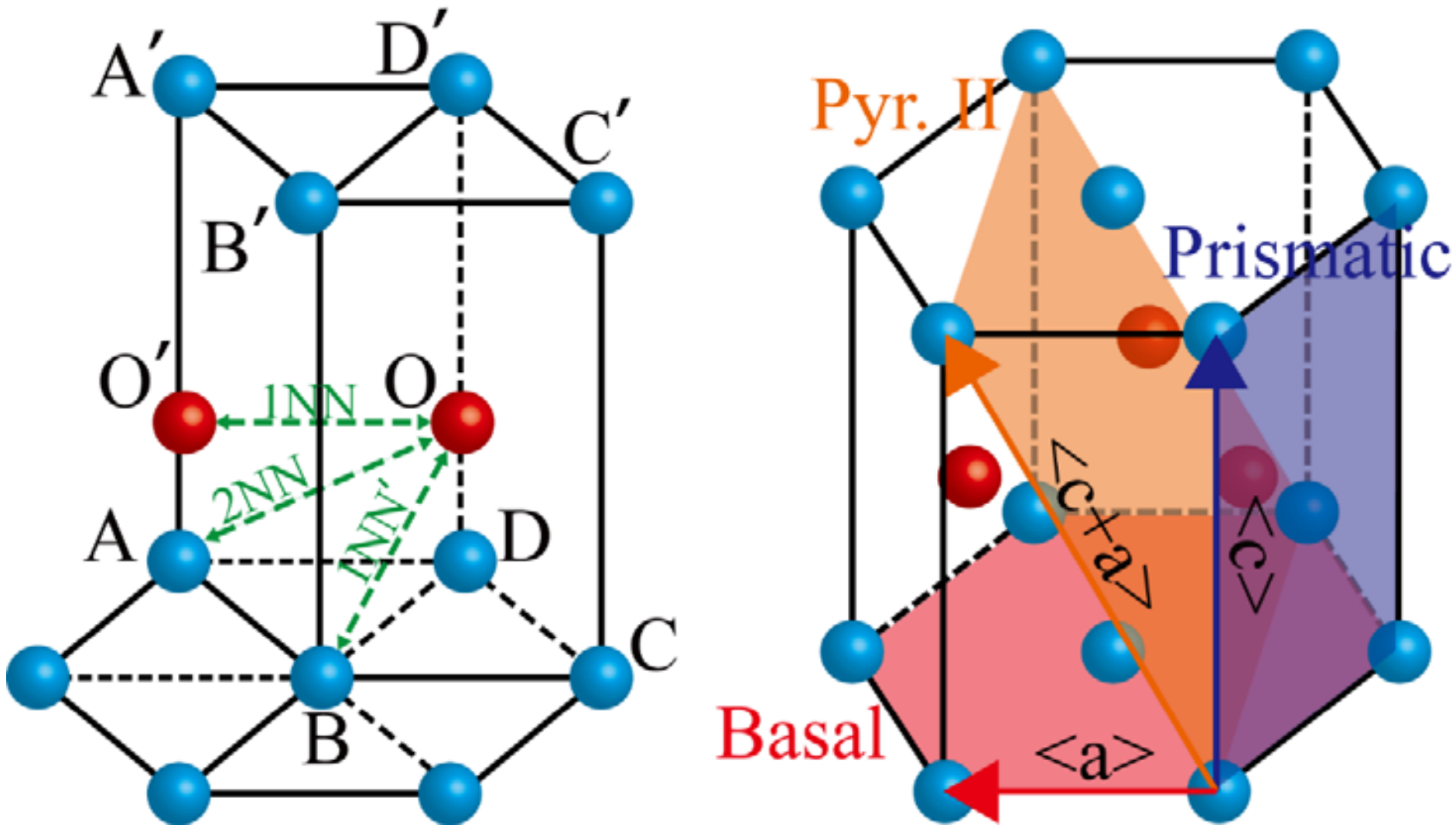

Figure 1 Schematic showing the (a) relative position of calculations the interaction between solutes and (b) the basal, prismatic and pyramidal II planes in a Mg unit cell.

The influence of solutes on unstable SFEs ($E_{usf}$) and stable SFEs ($E_{sf}$) of basal, prismatic, and Pyramidal II slips were investigated (see Fig. 1b). As illustrated in Fig. 2, the solute atoms may sit at different locations, rendering the solute atom pairs assuming various configurations with respect to the slip plane. For the 1NN, 1NN' and 2NN solute pairs we considered, the non-equivalent configurations were listed in Table 1. All the calculations were carried out using Vienna Ab-initio Simulation Package (VASP) have been performed on a series of ternary Mg alloys. The projector augmented wave method (PAW) was used, accompanying with Perdew–Burke–Ernzerhof (PBE) exchange-correlation functional for the generalized-gradient-approximation (GGA) [22-24]. A cut-off energy of 400 eV, and a 7×7×1 Monkhorst-Pack *k*-point mesh were used in our calculations. Internal coordinates of atoms within the supercells as well as the shape of the supercells were fully relaxed by the first-order Methfessel-Paxton smearing method with the width = 0.1 eV. Additional benchmark calculations were also performed to confirm that the dimensions of the supercell are sufficient. The energy and force convergence criteria were set as $10^{-5}$ eV and $10^{-2}$ eV/Å, respectively.

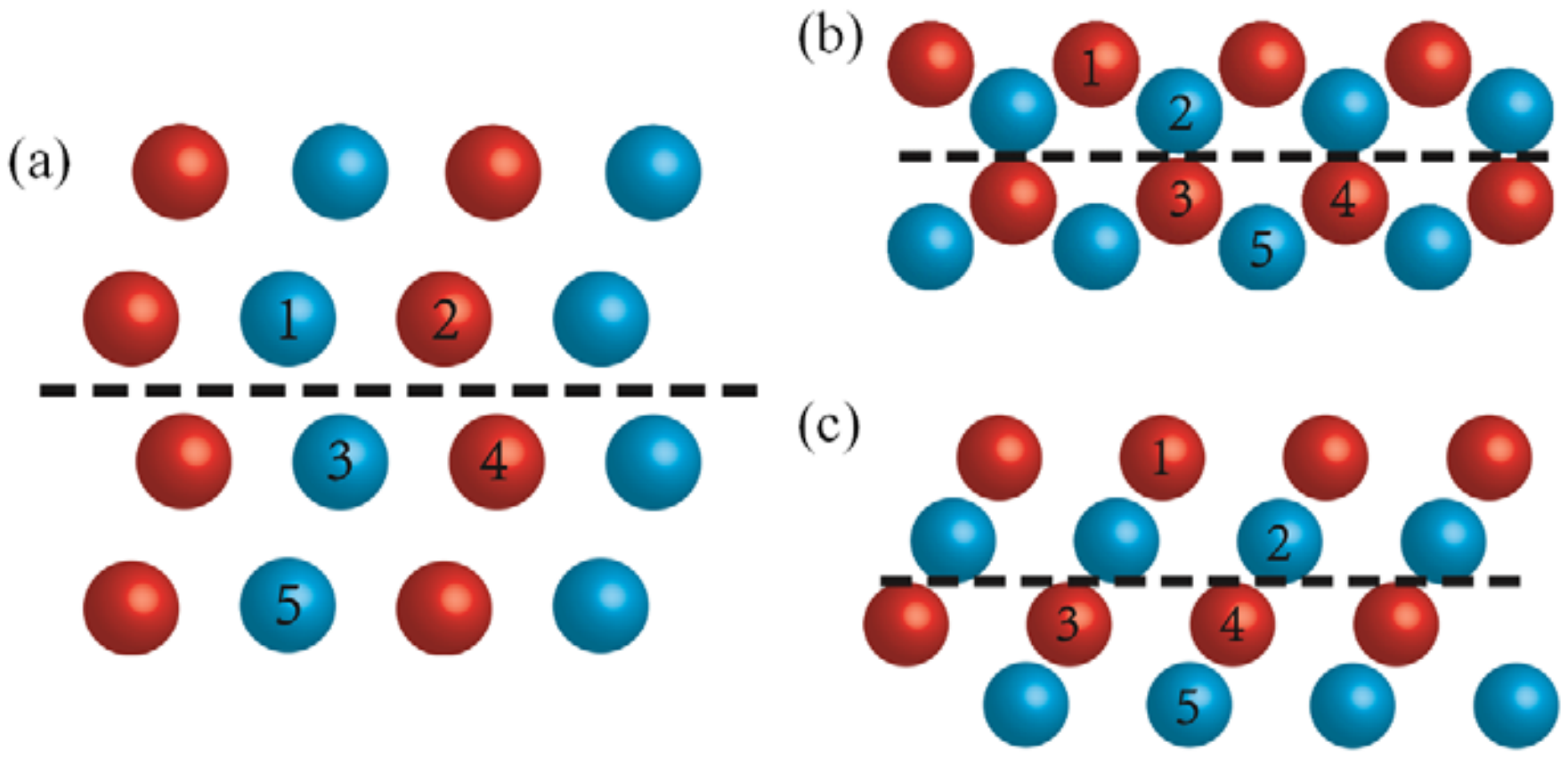

Figure 2 Configurations of site distributions of solute pairs in (a) basal slip (b) prismatic slip and (c) Pyramidal II slip.

Table 1 The list of non-equivalent configurations for solute pairs, with the sites (c.f., Fig. 2) assumed by the two solute atoms (denoted as X and Y) within the solute pair indicated.

| Slip system | Configuration | Solute X | Solute Y | Pair type |
|---|---|---|---|---|
| Basal | $\alpha_1$ | 3 | 4 | 1NN |
| | $\beta_1$ | 3 | 5 | 1NN' |
| | $\gamma_1$ | 1 | 3 | 1NN' |
| | $\delta_1$ | 2 | 3 | 1NN' |
| | $\varepsilon_1$ | 1 | 4 | 2NN |
| | $\epsilon_1$ | 4 | 5 | 2NN |
| Prismatic | $\alpha_2$ | 3 | 4 | 1NN |
| | $\beta_2$ | 3 | 5 | 1NN' |
| | $\gamma_2$ | 1 | 3 | 1NN' |
| | $\delta_2$ | 2 | 3 | 1NN' |
| | ε2 | 1 | 4 | 2NN |
| | ϵ2 | 4 | 5 | 2NN |
| Pyramidal | α3 | 3 | 4 | 1NN |
| | β3 | 3 | 5 | 1NN’ |
| | γ3 | 1 | 3 | 1NN’ |
| | δ3 | 2 | 3 | 1NN’ |
| | ε3 | 1 | 4 | 2NN |
| | ϵ3 | 4 | 5 | 2NN |

## 3. Results and Discussion

### 3. 1 Binding energy of solutes in Mg alloy

To determine the interaction between solutes, we examined the binding energy $E_b^{XY}$ between two solute atoms X and Y, calculated using the formula below:

$$E_b^{XY} = E[\mathrm{Mg}_{n-2}\mathrm{XY}] - E[\mathrm{Mg}_{n-1}\mathrm{X}] - E[\mathrm{Mg}_{n-1}\mathrm{Y}] + E[\mathrm{Mg}_n] \quad (1)$$

where $n$ is an integer denoting the total number of atoms within the supercell, $E[\mathrm{Mg}_{n-2}\mathrm{XY}]$ is the energy of the bulk with solute X and Y at first nearest-neighbors (1NN or 1NN’) or second nearest-neighbors (2NN), $E[\mathrm{Mg}_{n-1}\mathrm{X}]$ and $E[\mathrm{Mg}_{n-1}Y]$ are the energies of a bulk with a single solute X or Y, respectively, and $E[\mathrm{Mg}_n]$ is the energy of a pristine Mg bulk. A negative binding energy is indicative of attraction between the two solutes while a positive binding energy signals repulsion. Part of the calculated binding energy data are shown in Fig. 3, from which several observations can be made. First, we see that for a solute pair, the two constituting elements arranged as either the in-plane 1NN or out-of-plane1NN’, mostly exhibit similar binding energy values although exceptions exist. The similarity in binding energies for 1NN and 1NN’ comes from the sensitive to the distance between solutes not in or out of plane. Second, we can see a particular correlation between the binding energies of 1NN and 2NN. They are mostly of opposite sign, i.e., one would expect the binding energy of 2NN to be negative if the binding energy of 1NN is positive, and vice versa. Further examining the results from Fig. 3 with reference to the characteristics of the solute elements, listed in Table 2, we can see that the 1NN binding energy tends to exhibit a positive value if both solute atoms in the pair have larger radii than Mg (e.g., Ca-Nd, Ca-Gd, and Nd-Y), while the 2NN binding energy, on the contrary, would yield a negative value, indicative the solute pair as energetically unfavorable and favorable at 1NN and 2NN positions. This can be understood from the aspect of mechanical misfit. Both larger solutes in the pair bring high misfit volume and dilation strain. When the pair assumes a 1NN configuration where the solutes are within close vicinity to each other, the two dilation fields would have significant overlap, thus considerably raising the mechanical energy and the system energy. For the same reason, the solute pair with one larger solute and one smaller solute, exhibits opposite misfit volumes at the two solutes which may form a "complementary effect" [25], making the pair energetically favorable at 1NN. Nonetheless, we see that there are exceptions to those afore-mentioned general observations. For instance, Ca and Mn always repel each other, regardless of the pair configurations. This may come from the huge electronegativity difference between the two solutes. Taking Gd-Mn, Gd-Zn and Mn-Nd as examples, the binding energy of them are -0.18ev, -0.08ev, and -0.11ev, respectively. This means there are quite high attractions between these solutes. As a result, these solutes are very likely to form solute pairs or even clusters in Mg alloys. Their effects on SFE will be significantly different from that of a single solute and worthy to study.

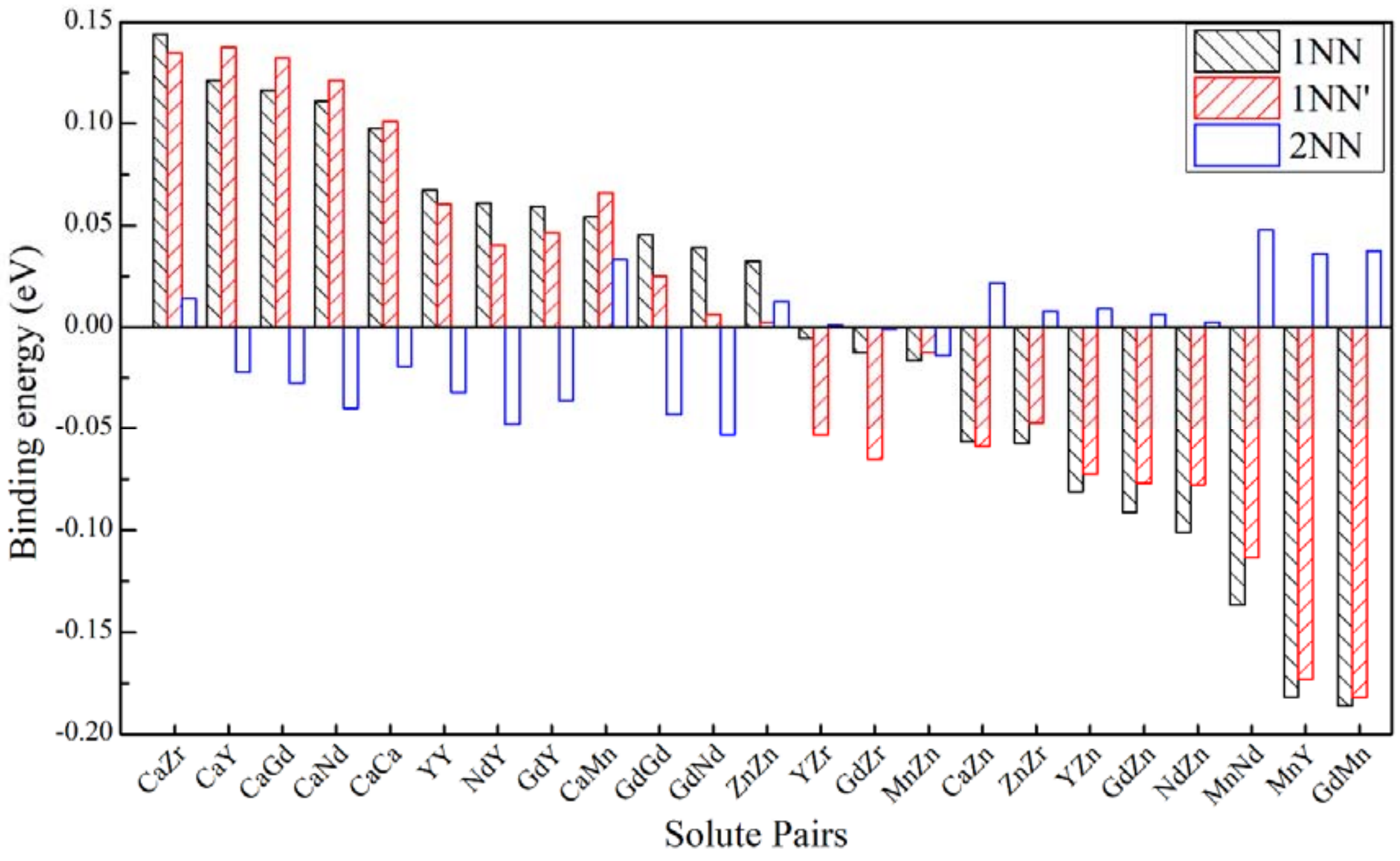

Figure 3 The binding energy of selected solute pairs.

Table 2 Atomic radius and electronegativity of Mg and the solutes considered in this study. $R_X$ and $R_{Mg}$ denote the atomic radii of the solute X and Mg respectively, while $Eln_X$ and $Eln_{Mg}$ are the electronegativities of the solute X and Mg respectively.

| Solute | Atomic radius (ai) | $\frac{R_X - R_{Mg}}{R_{Mg}}$ (%) | Electronegativity | $Eln_X - Eln_{Mg}$ |
|---|---|---|---|---|
| Mg | 160 | 0 | 1.31 | 0 |
| Ca | 197 | 23.13 | 1 | -0.27 |
| Gd | 180.4 | 12.75 | 1.2 | -0.09 |
| Mn | 140 | -12.5 | 1.55 | 0.17 |
| Nd | 181.4 | 13.19 | 1.14 | -0.14 |
| Y | 180 | 12.5 | 1.22 | -0.07 |
| Zn | 134 | -16.25 | 1.65 | 0.23 |
| Zr | 160 | 0 | 1.33 | 0.02 |

### 3. 2 Effect of alloying on SFE in ternary Mg alloys

According to the binding energy data obtained above, we note that there exist multiple alloy pair configurations yielding negative binding energies, indicative of attraction and these pairs being energetically favorable to form. Among these pairs, there are 1NN or 1NN' pairs, and 2NN pairs. However, for 2NN pairs, our preliminary calculations indicated that the influence trend of 2NN is almost identical to that of 1NN. Therefore, in our SFE calculations discussed in the follows, we only consider 1NN or 1NN' pairs. Also worth mentioning is that for the basal slip system, we examine both the stable and unstable stacking fault energies, i.e., basal $E_{sf}$, basal $E_{usf}$. Because stable stacking fault does not exist on prismatic plane [26]. And for pyramidal plane, as noted by Yin et al. [27], determining the $E_{sf}$ on the pyramidal plane solely from the local minimum of the GSFE curve is inadequate. Therefore, we only examine their unstable stacking fault energies.

We listed all the possible solute pairs that may formed between the 16 alloying elements. Next, we calculated the basal $E_{sf}$, basal $E_{usf}$, prismatic $E_{usf}$, and Pyramidal II $E_{usf}$ of Mg with these possible solute pairs. The results of these SFE under config α, in which the two solutes located in the same layer below the slip plane, were shown in Fig. 13. For basal SFE in Mg alloys, the solute pairs which decrease the $E_{sf}$ will also decrease the $E_{usf}$. Only a few solute pairs increase the SFE in basal plane. And most of them contain Mn, like Mn-Ti, Mn-Al, Mn-Dy, from which we can deduce that Mn is a good choice if higher SFE are preferred in alloy designing. When it comes to the prismatic slip plane, the $E_{usf}$ were decreased by up to 50% with RE-X solute pairs, such as Nd-Al, La-La. The situation for SFE in Pyramidal II becomes more complicated. More detailed analysis is necessary for a comprehensive understanding of the effect of solute pairs.

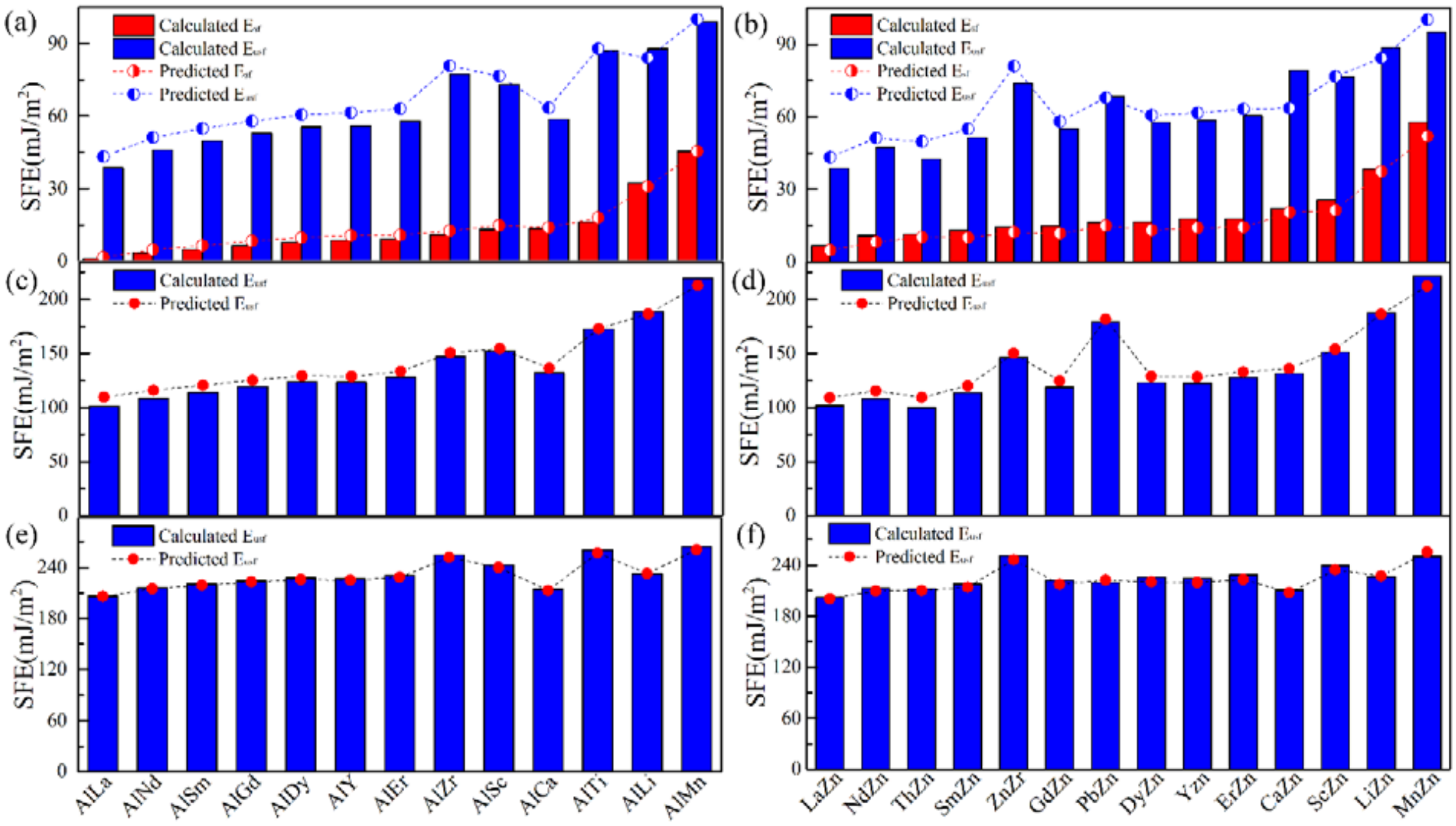

Figure 4 The calculated and predicted results of (a) Mg-Al-X basal, (b) Mg-Zn-X basal, (c) Mg-Al-X prismatic, (d) Mg-Zn-X prismatic, (e) Mg-Al-X pyramidal and, (f) Mg-Zn-X pyramidal SFE under config α

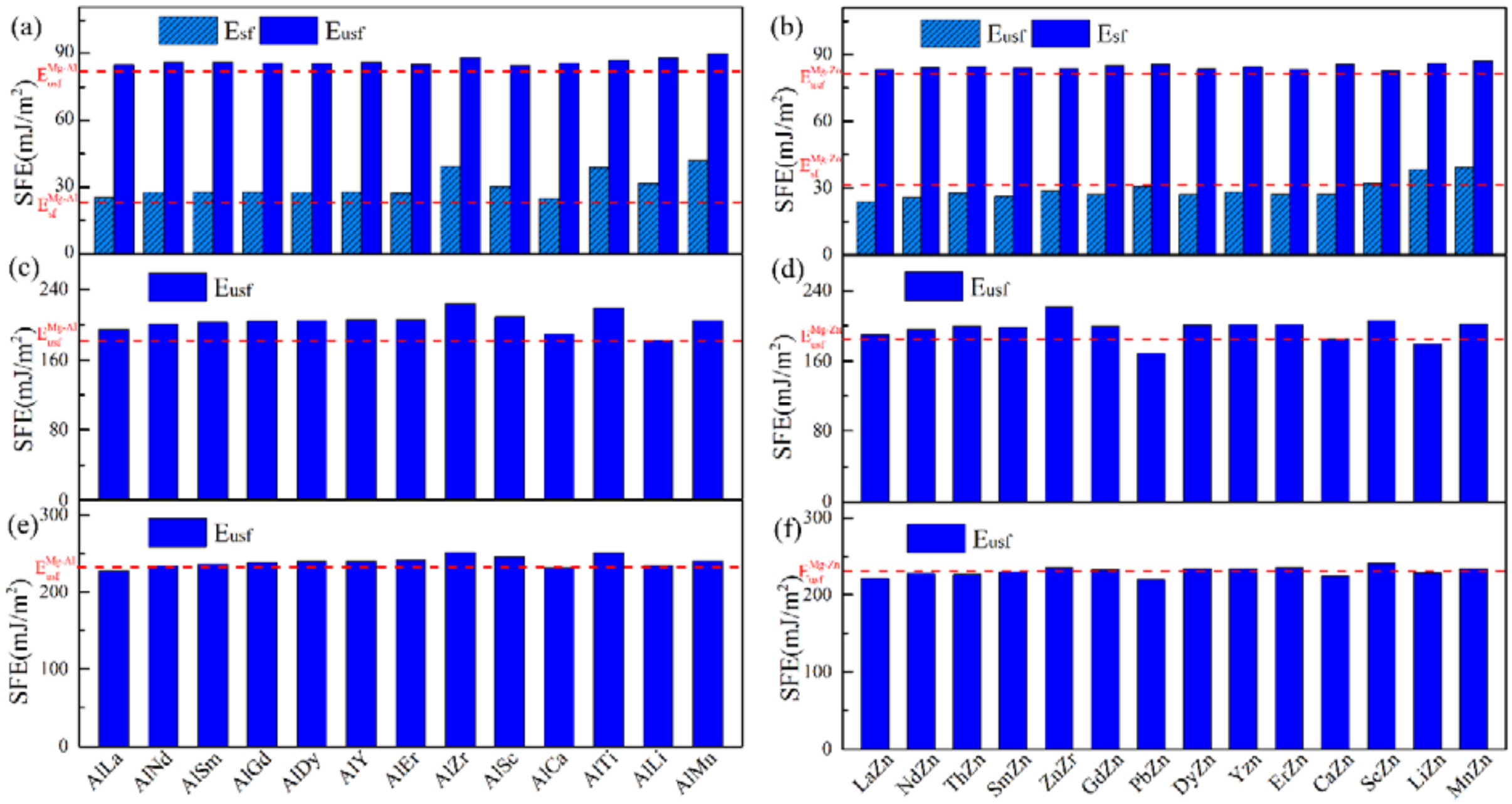

Figure 5 The calculated results of (a) Mg-Al-X basal, (b) Mg-Zn-X basal, (c) Mg-Al-X prismatic, (d) Mg-Zn-X prismatic, (e) Mg-Al-X pyramidal and, (f) Mg-Zn-X pyramidal SFE under config β

We here chose two series of typical Mg ternary alloys, Mg-Al-X and Mg-Zn-X, to further show the solute pair influence. The calculated SFE of the model with Al-X and Zn-X solute pairs under config α are shown in Fig. 4. As mentioned before, the two solutes were at the same layer near the slip plane. Since the two solutes located at the equivalent sites, we conjecture that the impact of the solute pair on SFE in this circumstance may be approximately equal to the combination of the impact of each single solute. Accordingly, the following equation was used to predict the effect of the solute pairs by adding the effect of each single solute and the SFE of pure Mg.

$$E_{\alpha}^{Mg-X-Y} = E_{sf}^{Mg} + (E_{sf}^{Mg-X} - E_{sf}^{Mg}) + (E_{sf}^{Mg-Y} - E_{sf}^{Mg}) = E_{sf}^{Mg-X} + E_{sf}^{Mg-Y} - E_{sf}^{Mg} \quad (2)$$

where $E_{\alpha}^{Mg-X-Y}$ is the SFE of $Mg_{N-2}XY$ system under config α, $E_{sf}^{Mg}$ is the basal stable SFE of pure Mg, $E_{sf}^{Mg-X}$ and $E_{sf}^{Mg-Y}$ are the basal stable SFE of Mg-X and Mg-Y system, respectively. The effect of the single solute was from our previous research [25]. The prediction results were shown as the points in Fig. 4. Good agreement was achieved for both Mg-Al-X and Mg-Zn-X. The results of Mg-Mn-X in Fig. 14 also proved our hypothesis.

As for the config β, the two solutes located at different layer. According to Yin et al.'s study on solute interaction with SF, the effect of solute that near the stacking fault layer play the dominant role. In this model, Al or Zn was near the slip plane. So, the effect of the solutes at second-near layer was minor. Especially for the $E_{usf}$, it can be seen in Fig. 5 and Fig. 14 that the value of $E_{usf}$ remain nearly unchanged with the solute pairs. $E_{sf}$ slightly changed with the solute pairs and this change was corresponded with the single solute effect. Therefore, we can roughly use $E_{\beta}^{Mg-X-Y} = E_{sf}^{Mg-X}$ to predict the SFE under config β. Or more accurately for $E_{sf}$ we have:

$$E_{\beta}^{Mg-X-Y} = E_{sf}^{Mg-X} + \theta * (E_{sf}^{Mg-Y} - E_{sf}^{Mg}) \quad (3)$$

where X is the solute near the slip plane, and Y is the one at second-near layer, $\theta$ is the corresponding coefficient obtained by linear fitting.

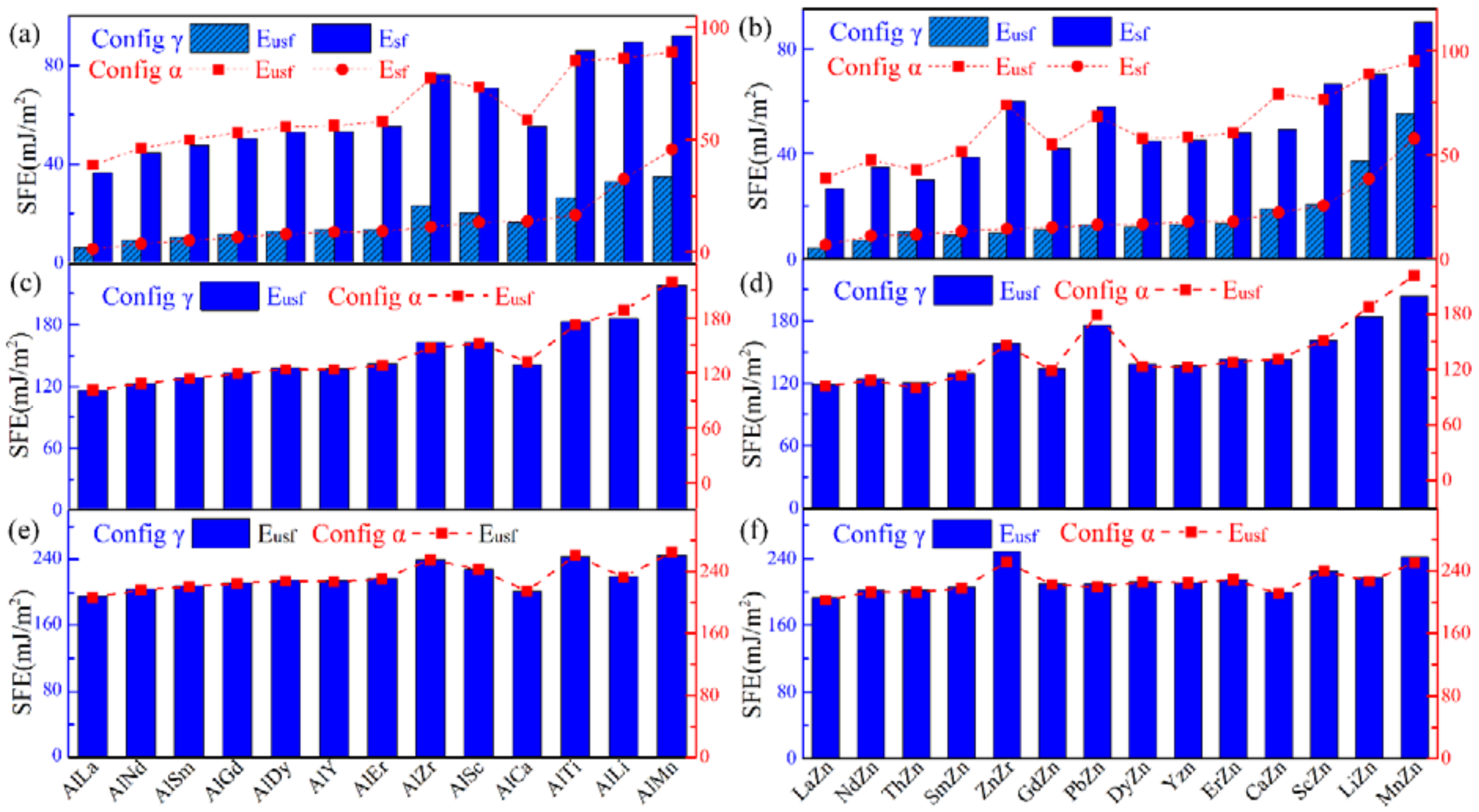

Figure 6 The calculated results of (a) Mg-Al-X basal, (b) Mg-Zn-X basal, (c) Mg-Al-X prismatic, (d) Mg-Zn-X prismatic, (e) Mg-Al-X pyramidal and, (f) Mg-Zn-X pyramidal SFE under config α and γ

Fig. 6 shows the calculation results of the $E_{sf}$ and $E_{usf}$ under config α and γ. To enhance the clarity of the observed trends, a modified scale was employed for the SFE under configuration α. In general, the SFE values under configuration α exceed those under configuration γ. With the exception of certain solute pairs such as Ca-Zn and Al-Zr, the influence trends of solute pairs remain consistent between configurations α and γ. Consequently, we can utilize the same approach as outlined in equation (2) to forecast the SFE under configuration γ. This similarity arises from the fact that both solutes are situated in close proximity to the slip plane. In configuration α, solutes are positioned below the slip plane, whereas in configuration γ, solutes traverse the slip plane.

Another intriguing thing was found when we try to predict the SFE under config δ using machine learning (ML). The SFE shows a strong correlation with binding energies in the joint distribution of the training set. With the help of ML, the SFE under config δ can be derived by $E_{\alpha}^{\mathrm{Mg-X-Y}}$:

$$E_{\delta}^{\mathrm{Mg-X-Y}} = E_{\alpha}^{\mathrm{Mg-X-Y}} + \mu * (E_{b-1NN} - E_{b-2NN}) \quad (5)$$

where $E_{\delta}^{\mathrm{Mg-X-Y}}$ and $E_{\alpha}^{\mathrm{Mg-X-Y}}$ are the SFEs with solute pairs under config δ and α, respectively. $E_{b-1NN}$ and $E_{b-2NN}$ denote the binding energy of the solute pairs at 1NN and 2NN, respectively. In this study, basic regression techniques were employed to determine the value of μ. The anticipated outcomes of the SFE for all slip systems, as calculated using eq. (5), are presented in Fig. 7. The graphical representation demonstrates that the model's predictions align reasonably well with the DFT results. The mechanism underlying this model's performance in basal slip can be elucidated as follows: Initially, the solutes are positioned in the 1NN configuration, as illustrated in Fig. 15. In the stable stacking fault configuration, the distance between solute pairs is maintained at 4.50 Å, mirroring that of the 2NN configuration. Based on the configuration disparity, we postulate that the SFE is connected to the difference in binding energy between the 1NN and 2NN configurations.

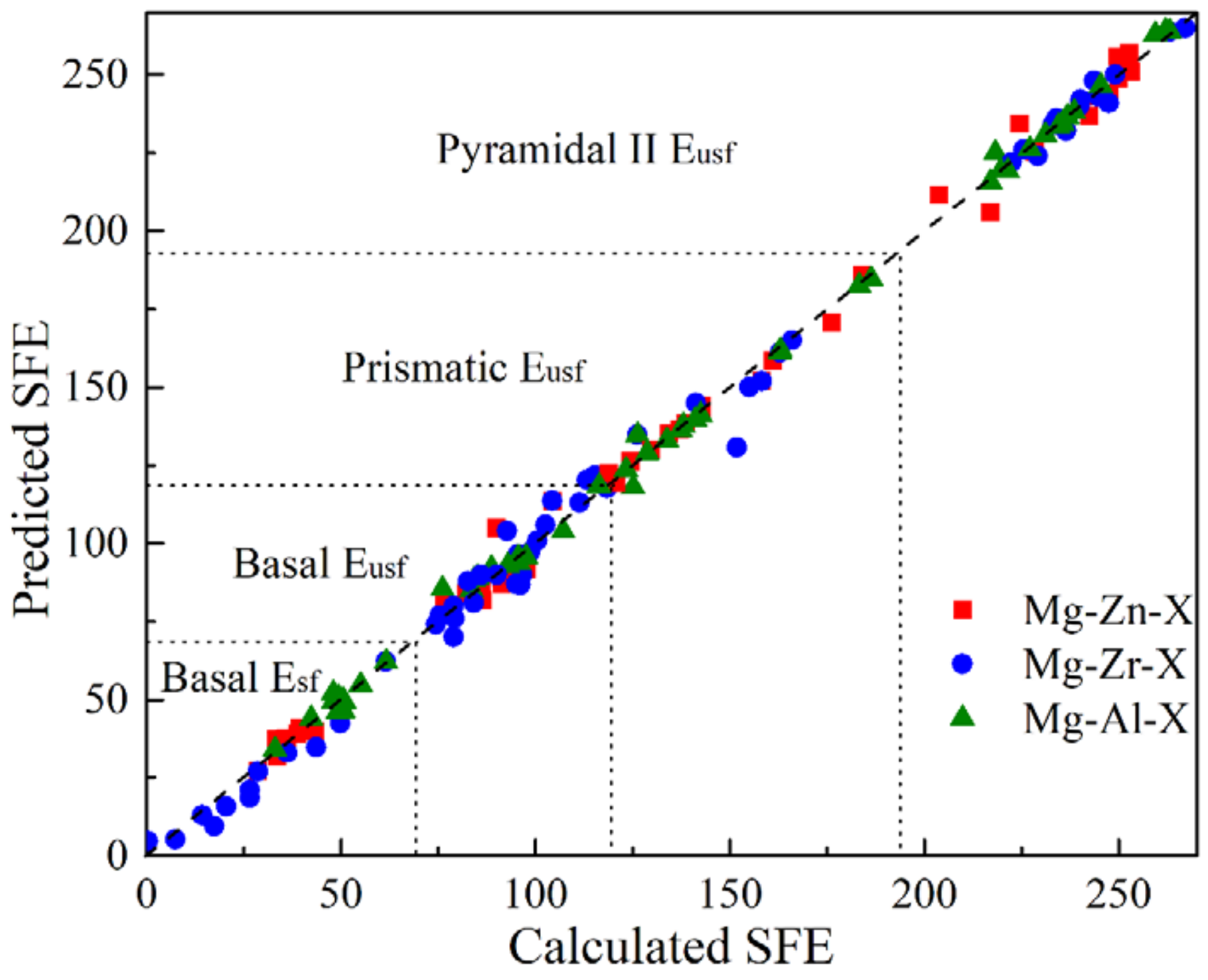

Figure 7 The calculated and predicted $E_{sf}$ and $E_{usf}$ under config δ.

### 3.3 Potential Mg alloy designing solute pair candidates

As mentioned at beginning, RE elements are beneficial to the ductility of Mg alloys. But the mechanism behind them is still under debating. More research is needed to have a better understanding of the RE element behavior during the deformation process. That is also why we performed the comprehensive study on SFE in ternary Mg alloys. Inspired by the idea of machine learning, there should be an approach to achieve high ductility Mg alloy designing without fully understanding the deformation mechanism. Since RE elements are favorable for deformation and SFE is a crucial parameter control the deformation. We came up with an idea that the solute pairs with similar SFE to RE elements may also have excellent ductility. To seek the potential solute pairs, the following equation was used to determine their similarity:

$$\gamma_D = \sum_{i=0}^{3}\left(\frac{\gamma_i^{Mg-X-Y} - \gamma_i^{Mg-RE}}{\gamma_i^{Mg}}\right)^2 \tag{6}$$

where $\gamma_i^{Mg-X-Y}$ and $\gamma_i^{Mg-RE}$ are the SFE of Mg-X-Y and Mg-RE system, respectively. I=0, 1, 2, 3 refer to the basal stable SFE, basal unstable SFE, prismatic unstable SFE, and Pyramidal II unstable SFE, respectively. The smaller the $\gamma_D$ is, the effect of solute pairs is closer to that of RE elements. After calculating all the possible solute pairs considered in this study, the 7 pairs possess the lowest $\gamma_D$ were shown in Fig. 8. Among these, some compositions have already been proved with high ductility, such as Mg-Ca-Zn and Mg-Sn-Zr [26, 27]. We believe these solute pair candidates will bring new inspiration to Mg alloy design.

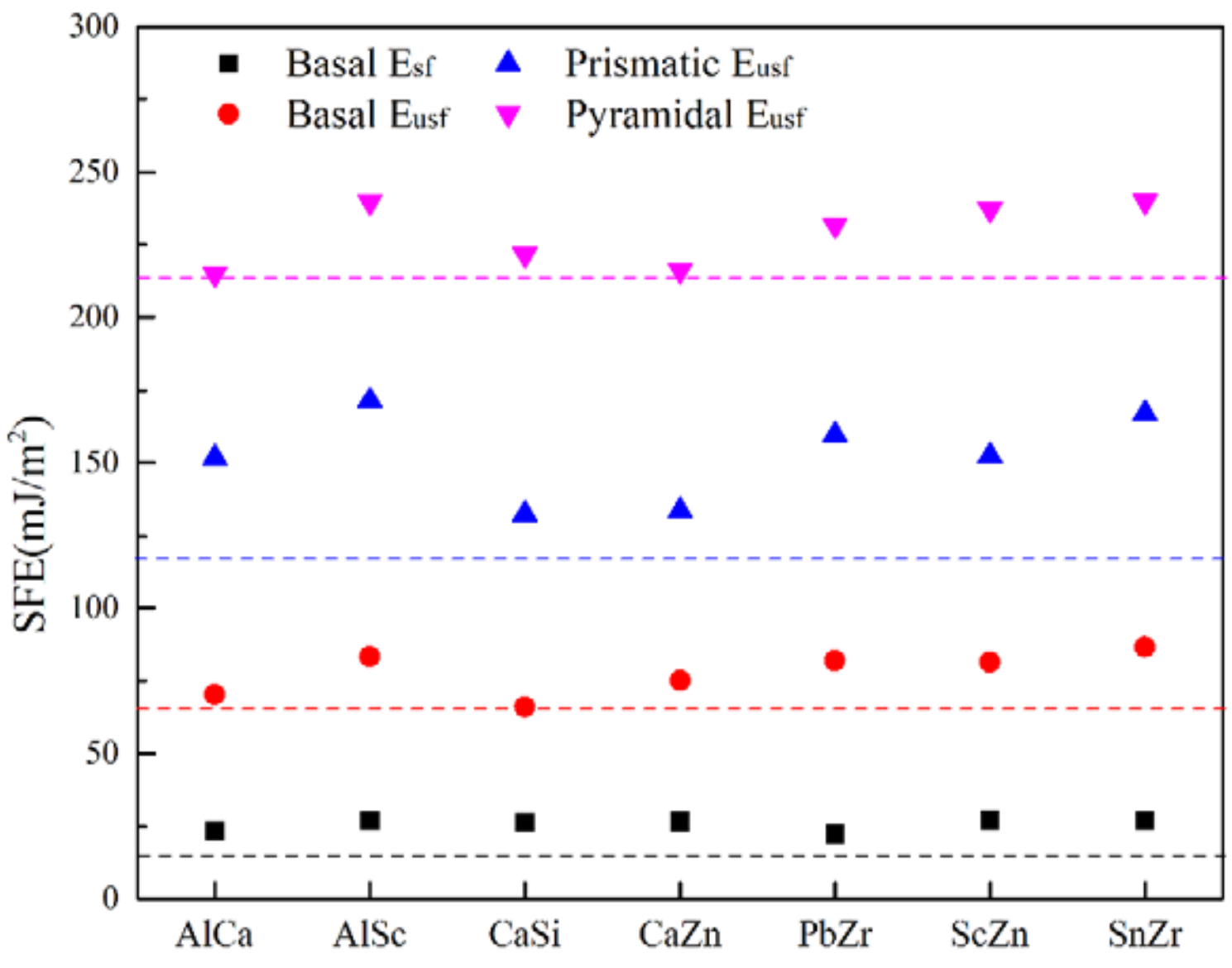

Figure 8 The non-RE solute pairs with lowest $\gamma_D$ and their SFEs.

As mentioned in section 1, limited number of independent slip systems in hcp mainly cause the low formability of Mg alloys at room temperature. Therefore, it's necessary to active non basal slip to achieve better plastic deformation. To activate prismatic or pyramidal slip systems, it essentially requires reducing their unstable SFE on the slip plane. The average reduction effect Δ of solute pairs on SFE under all configuration is defined as:

$$\Delta = 1/n \sum_{i=0}^{n} (E_{usf}^{Mg} - E_{usf}^{Mg-X-Y})/E_{usf}^{Mg} \quad (6)$$

where n is the total number of the config, $E_{usf}^{Mg}$ and $E_{usf}^{Mg-X-Y}$ are the unstable SFE of pure Mg and Mg-X-Y system. Then, we have

$$\Lambda_{p/b} = \frac{\Delta_{prismatic}}{\Delta_{basal}} \text{ or } \frac{\Delta_{pyramidal}}{\Delta_{basal}} \quad (7)$$

when $\Lambda_{p/b} > 1$, it means the solute pair has stronger reduction effect on non-basal SFE than that of basal SFE. The results of $\Lambda_{p/b}$ of all the solute pairs that can reduce the non-basal SFE were shown in Fig. 9. Logarithm value was taken to show the results more intuitively in Fig. 9a.

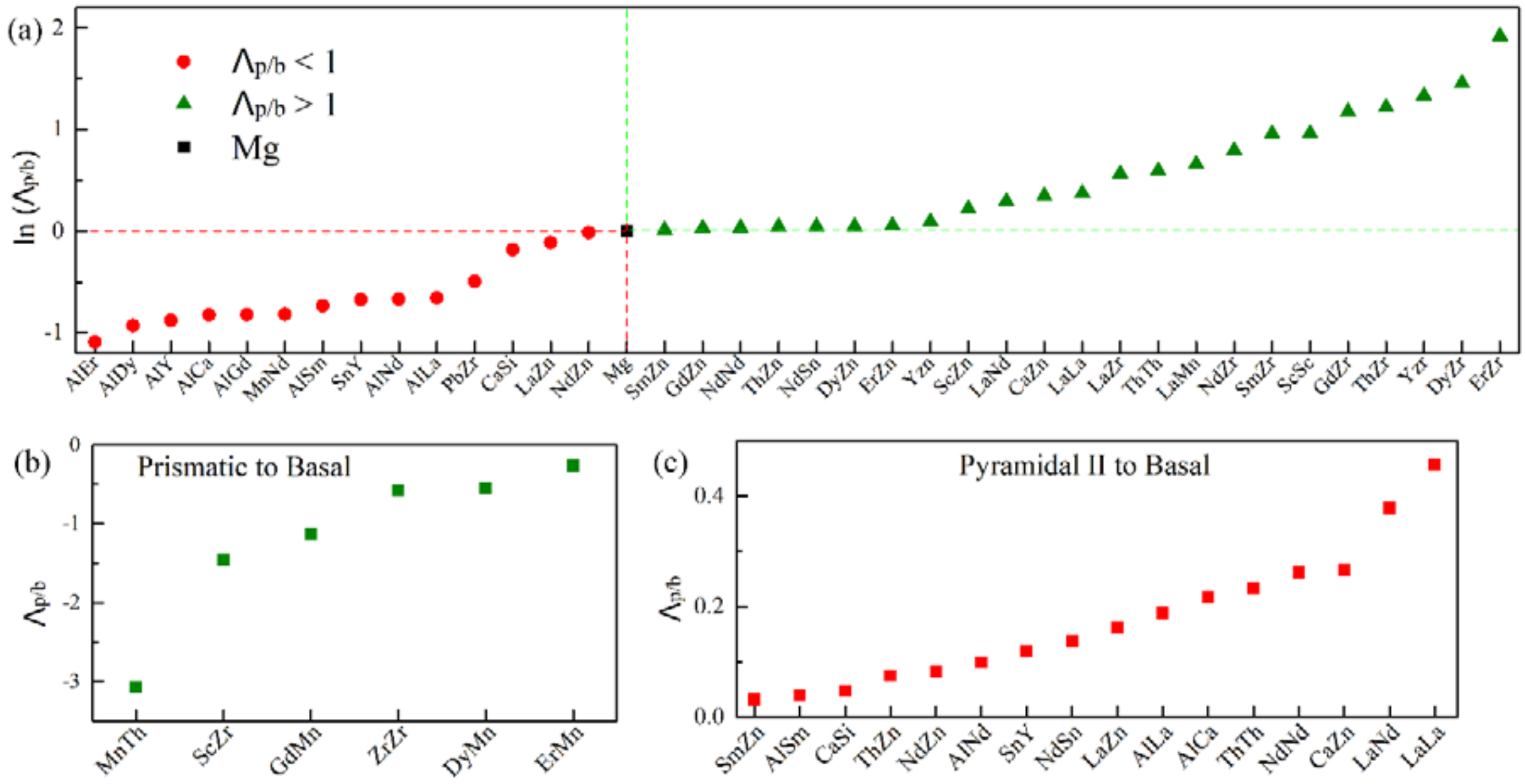

Figure 9 (a) The logarithm value of $\Lambda_{\frac{prismatic}{basal}}$ (b) $\Lambda_{\frac{prismatic}{basal}}$ of selected solute pairs (c) $\Lambda_{\frac{pyramidal}{basal}}$ of selected solute pairs

Al-X series solute pairs all distributed in the red zone, which means they have strong reduction effect on basal SFE. The effect of Zn-X on basal and prismatic SFE is nearly the same since the value of $\frac{\Delta_{prismatic}}{\Delta_{basal}}$ is around 1. Zr-X solute pairs fell in the rightmost of figure. They significantly decrease the prismatic unstable SFE, thus making the activation of prismatic slip easier. It is worth noting that the values of $\Delta_{prismatic}$ to $\Delta_{basal}$ of some solute pairs are minus (shown in Fig. 9b). It indicated that these solute pairs would decrease the prismatic SFE and increase the basal SFE at the same time. It further lowers the anisotropy and makes benefit in ductility improvements. Only few solute pairs (mainly RE-X pairs) reduce the pyramidal unstable SFE. And the ratio of $\Delta_{pyramidal}$ to $\Delta_{basal}$ shows that the reduction effect was not obvious on pyramidal slip plane. Therefore, lowering the activation energy of pyramidal slip may not be the reason of improvement of ductility by RE elements.

These results can explain some phenomena in our experiments. Strong basal texture has always been found in pure Mg and Mg alloys due to the extremely low basal slip activation energy [29-32]. For other hcp metals with c/a ratio $< 1.633$, such as titanium (Ti), have relative lower prismatic SFE. Then the TD-split texture was commonly observed in Ti deformation. While the characterization upon annealing revealed that the TD-split texture was also generated in ZEK100 (nominal composition: Mg 98.8 Zn 1 Zr 0.1 RE 0.1) and ZE02 (nominal composition: Mg 99.8 Nd 0.2). The RE element or the favorable Zr-RE solute pairs decrease the prismatic SFE and promote the basal poles splitting toward the TD.

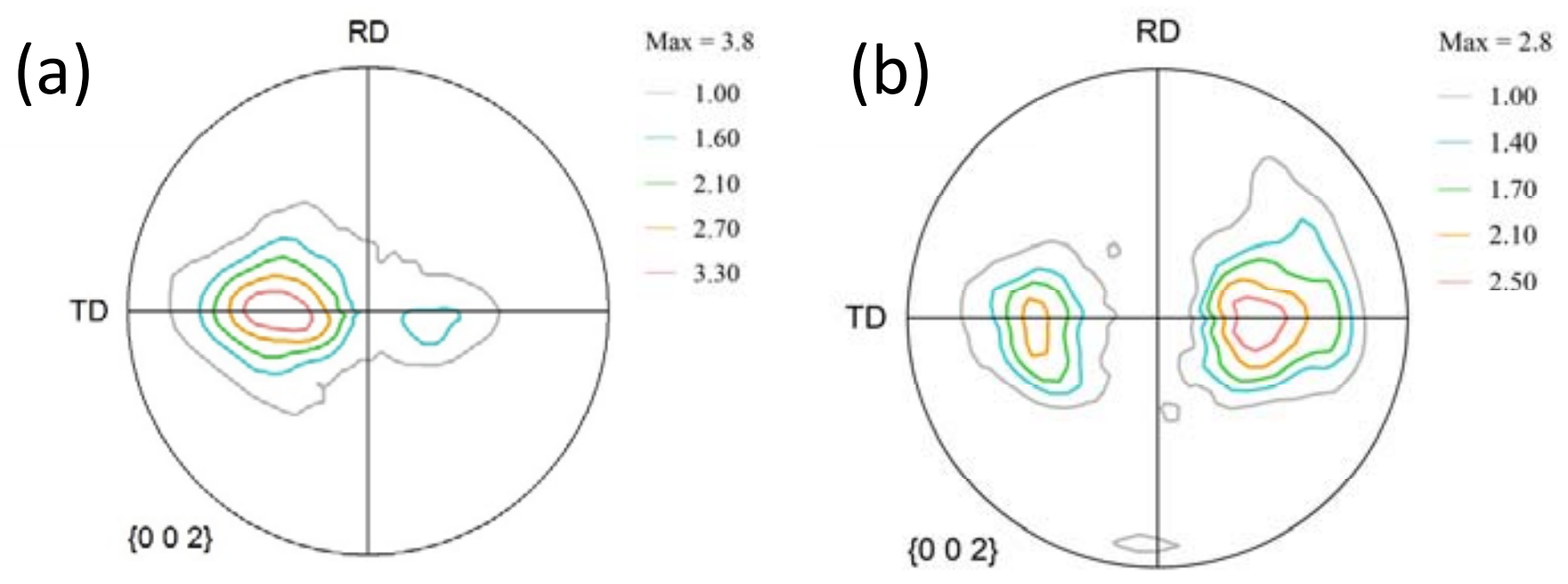

Figure 10 Macro-texture of fully annealed 550°C-solutionized (a) ZEK100 and (b) ZE-02

## 4 Conclusion

Our study delved into the binding energy between solutes, revealing intricate interactions that play a crucial role in the formation of solute pairs. This understanding is pivotal for anticipating the behavior of alloying elements in Mg alloys and their impact on SFE. Furthermore, the investigation extended to ternary Mg alloys, particularly focusing on Mg-Al-X and Mg-Zn-X systems. Predictive models were proposed to estimate the effect of solute pairs on SFE under different configurations, providing practical guidelines for alloy design. The study identified several solute pairs with a strong potential to influence ductility in Mg alloys. By assessing the similarity of SFE patterns to RE elements, we pinpointed seven promising solute pairs. These candidates open avenues for novel inspirations in Mg alloy design, aligning with the pursuit of enhanced ductility.

In conclusion, this research not only contributes to the fundamental understanding of SFE in Mg alloys but also offers valuable insights for the development of advanced Mg alloys with improved mechanical properties.

## Acknowledgments

We gratefully acknowledge the financial support from the Natural Sciences and Engineering Research Council of Canada (NSERC) Discovery grant (RGPIN-2017-05187) and China Scholarship Council (CSC). We also acknowledge Compute Canada for providing computing resources.

## References

[1] Z. Wu, W. Curtin, The origins of high hardening and low ductility in magnesium, Nature 526 (2015) 62-67.

[2] T.M. Pollock, Weight loss with magnesium alloys, Science 328 (2010) 986-987.

[3] G. G. Yapici, S. V. Sajadifar, A. Hosseinzadeh, T. Wegener, C. Sobrero, A. Engelhardt, T. Niendorf, Effect of Friction Stir Processing on the Fatigue Performance of AZ31 Magnesium Alloy, Adv. Eng. Mater. (2023) 2201638.

[4] S.R. Agnew, J.F. Nie, Preface to the viewpoint set on: the current state of magnesium alloy science and technology, Scr. Mater. 63 (2010) 671-673.

[5] C.M. Byer, K.T. Ramesh, Effects of the initial dislocation density on size effects in single-crystal magnesium, Acta Mater. 61 (2013) 3808-3818.

[6] Z. Wu, R. Ahmad, B. Yin, S. Sandlöbes, W.A. Curtin, Mechanistic origin and prediction of enhanced ductility in magnesium alloys, Science 359 (2018) 447-452.

[7] J.P. Hirth, J. Lothe, Theory of dislocations, 2 ed., John Wiley & Sons (1982).

[8] U. M. Chaudry, S. Tekumalla, M. Gupta, T. Jun, K. Hamad, Designing highly ductile magnesium alloys: current status and future challenges, Crit. Rev. Solid State Mater. Sci. 47 (2022) 194-281.

[9] A. P. Carvalho, R. B. Figueiredo, The Effect of Ultragrain Refinement on the Strength and Strain Rate Sensitivity of a ZK60 Magnesium Alloy, Adv. Eng. Mater. 24 (2022) 2100846.

[10] N. Zhou, Z. Zhang, L. Jin, J. Dong, B. Chen, W. Ding, The role of Ca, Al and Zn on room temperature ductility and grain boundary cohesion of magnesium, J. Magnes. Alloy, 9 (2021) 1521-1536.

[11] W.J. Kim, S.I. Hong, Y.H. Ki, Enhancement of the strain hardening ability in ultrafine grained Mg alloys with high strength, Scr. Mater. 67 (2012) 689-692.

[12] Y. Chino, K. Sassa, M. Mabuchi, Enhancement of tensile ductility of magnesium alloy produced by torsion extrusion, Scr. Mater. 59 (2008) 399-402.

[13] J. He, B. Jiang, Q. Yang, Y. Zeng, X. Xia, F. Pan, Improved ductility of magnesium alloy sheets by pre-hardening and annealing, Mater. Sci. Technol. 31 (2015) 1383-1387.

[14] K. Edalati, E. Akiba, W. J. Botta, Impact of severe plastic deformation on kinetics and thermodynamics of hydrogen storage in magnesium and its alloys, J. Mater. Sci. Technol. 146 (2023) 221-23.

[15] V. Vitek, Intrinsic stacking faults in body-centred cubic crystals, Philos. Mag. 18(154) (1968) 773-786.

[16] J.R. RICE, Dislocation nucleation from a crack tip An analysis based on the Peierls concept, J. Mech. Phys. Solids 40(2) (1992) 239-70.

[17] V. Vitek, Structure of dislocation cores in metallic materials and its impact on their plastic behaviour, Prog. in Mater. Sci. 36 (1992) 1-27.

[18] B. Yin, Z. Wu, W.A. Curtin, First-principles calculations of stacking fault energies in Mg-Y, Mg-Al and Mg-Zn alloys and implications for <c+a> activity, Acta Mater. 136 (2017) 249-261.

[19] Y. Wang, Z. Jia, J. Ji, S. Li, D. Liu, Evolution of Stacking Fault and Dislocation during Dynamic Recrystallization of Inconel 625 Alloy, Adv. Eng. Mater. 24 (2022) 2200657.

[20] S. Sandlöbes, M. Friák, S. Korte-Kerzel1, Z. Pei, J. Neugebauer, D. Raabe, A rare-earth free magnesium alloy with improved intrinsic ductility, Sci. Rep. 7 (2017) 10458.

[21] T. Yonezawa, K. Suzuki, S. Ooki, A. Hashimoto, The effect of chemical composition and heat treatment conditions on stacking fault energy for Fe-Cr-Ni austenitic stainless steel, Metall. Mater. Trans. A Phys. Metall. Mater. Sci. 44A (2013) 5884-5896.

[22] S. Shi, L. Zhu, H. Zhang, Z. Sun, R. Ahuja, Mapping the relationship among composition, stacking fault energy and ductility in Nb alloys: A first-principles study, Acta Mater. 144 (2018) 853-861.

[23] M. Muzy, Z. Pakieła, K.J. Kurzydłowski, Generalized stacking fault energies of aluminum alloys-density functional theory calculations, Metals. 8(10) (2018) 823-832.

[24] S.L. Shang, W.Y. Wang, B.C. Zhou, Y. Wang, K.A. Darling, L.J. Kecskes, S.N. Mathaudhu, Z.K. Liu, Generalized stacking fault energy, ideal strength and twinnability of dilute Mg-based alloys: A first-principles study of shear deformation, Acta Mater. 67 (2014) 168-180.

[25] Q. Dong, Z. Luo, H. Zhu, L. Wang, T. Ying, Z. Jin, D. Li, W. Ding, X. Zeng, Basal-plane stacking-fault energies of Mg alloys: A first-principles study of metallic alloying effects, J. Mater. Sci. Technol. 34 (2018) 1773–1780.

[26] S. Sandlöbes, Z. Pei, M. Friák, L.F. Zhu, F. Wang, S. Zaefferer, D. Raabe, J. Neugebauer, Ductility improvement of Mg alloys by solid solution: Ab initio modeling, synthesis and mechanical properties, Acta Mater. 70 (2014) 92-104.

[27] C. Wang, H.Y. Zhang, H.Y. Wang, G.J. Liu, Q.C. Jiang, Effects of doping atoms on the generalized stacking-fault energies of Mg alloys from first-principles calculations, Scr. Mater. 69 (2013) 445-448.

# Supplementary Information

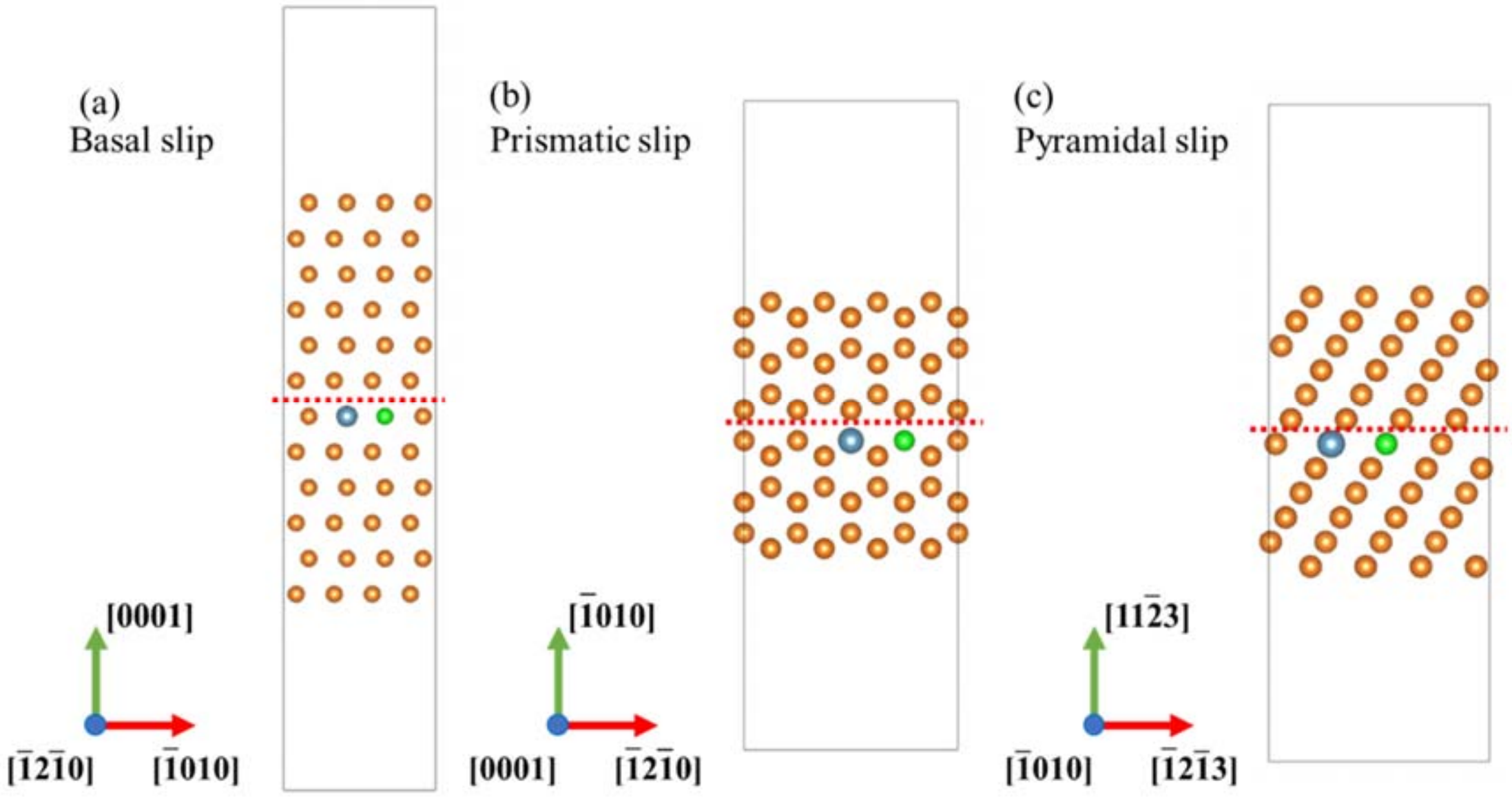

Figure 11 Supercells used in calculating SFE of (a) basal, (b) prismatic, (c) pyramidal plane (The orange spheres are Mg atoms, the blue and grey spheres are respectively solute atoms X and Y)

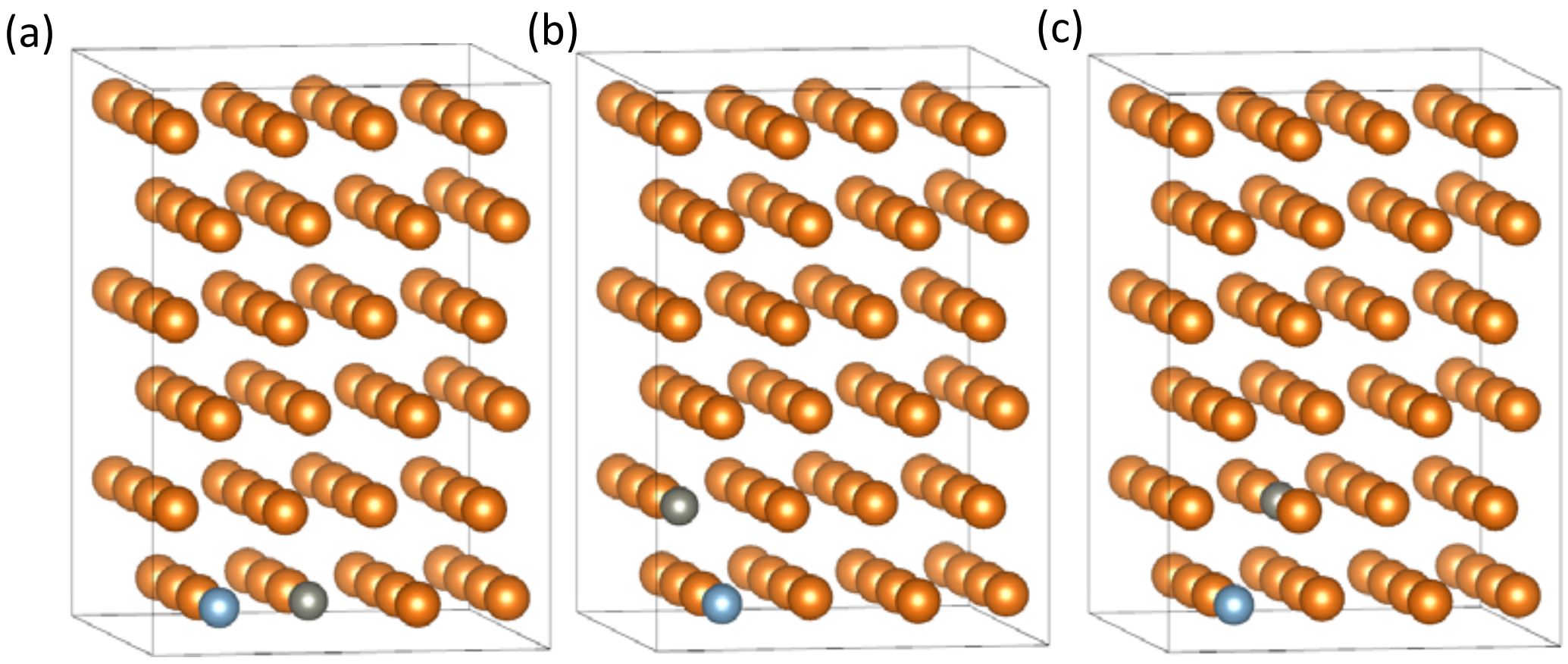

Figure 12 Supercells used in calculating binding energy with solutes locate at (a) 1NN, (b) 1NN’, (c) 2NN (The orange spheres are Mg atoms, the blue and grey spheres are respectively solute atoms X and Y)

Table 3 Binding energies of selected solute pairs at 1NN and 2NN location

| Solute Pairs | $E_b$ 1NN | $E_b$ 2NN | Solute Pairs | $E_b$ 1NN | $E_b$ 2NN | Solute Pairs | $E_b$ 1NN | $E_b$ 2NN |
|---|---|---|---|---|---|---|---|---|
| **Al-Al** | 0.02 | -0.03 | **Er-Er** | 0.01 | -0.13 | **Mn-Th** | -0.60 | 0.08 |
| **Al-Ca** | -0.07 | 0.06 | **Er-Gd** | 0.02 | -0.15 | **Mn-Ti** | -0.86 | -0.03 |
| **Al-Dy** | -0.13 | 0.06 | **Er-La** | -0.02 | -0.18 | **Mn-Y** | -0.17 | 0.03 |
| **Al-Er** | -0.12 | 0.06 | **Er-Mn** | -0.23 | 0.02 | **Mn-Zn** | -0.01 | -0.06 |
| **Al-Gd** | -0.13 | 0.06 | **Er-Nd** | 0.01 | -0.16 | **Mn-Zr** | -0.56 | 0.00 |
| **Al-La** | -0.14 | 0.08 | **Er-Sc** | 0.01 | -0.11 | **Nd-Nd** | -0.03 | -0.20 |
| **Al-Li** | -0.02 | 0.00 | **Er-Sm** | 0.02 | -0.16 | **Nd-Sc** | 0.02 | -0.13 |
| **Al-Mn** | -0.15 | -0.04 | **Er-Y** | 0.04 | -0.15 | **Nd-Sm** | -0.01 | -0.19 |
| **Al-Nd** | -0.13 | 0.07 | **Er-Zn** | -0.07 | 0.06 | **Nd-Sn** | -0.18 | 0.05 |
| **Al-Sc** | -0.11 | 0.04 | **Er-Zr** | -0.09 | -0.14 | **Nd-Y** | 0.04 | -0.18 |
| **Al-Si** | 0.03 | -0.04 | **Gd-Gd** | 0.02 | -0.17 | **Nd-Zn** | -0.08 | 0.08 |
| **Al-Sm** | -0.13 | 0.07 | **Gd-La** | -0.06 | -0.19 | **Nd-Zr** | -0.05 | -0.16 |
| **Al-Ti** | -0.15 | 0.00 | **Gd-Mn** | -0.18 | 0.03 | **Pb-Pb** | 0.07 | -0.02 |
| **Al-Y** | -0.12 | 0.06 | **Gd-Nd** | 0.01 | -0.18 | **Pb-Zn** | 0.04 | 0.00 |
| **Al-Zn** | 0.02 | -0.03 | **Gd-Sc** | 0.02 | -0.12 | **Pb-Zr** | -0.01 | 0.01 |
| **Al-Zr** | -0.17 | 0.02 | **Gd-Sm** | 0.02 | -0.18 | **Sc-Sc** | -0.01 | -0.09 |
| **Ca-Ca** | 0.10 | -0.10 | **Gd-Sr** | 0.13 | -0.16 | **Sc-Sm** | 0.02 | -0.12 |
| **Ca-Dy** | 0.13 | -0.12 | **Gd-Y** | 0.05 | -0.17 | **Sc-Y** | 0.02 | -0.12 |
| **Ca-Er** | 0.13 | -0.11 | **Gd-Zn** | -0.08 | 0.07 | **Sc-Zn** | -0.05 | 0.03 |
| **Ca-Gd** | 0.13 | -0.12 | **Gd-Zr** | -0.07 | -0.15 | **Sc-Zr** | -0.12 | -0.12 |
| **Ca-La** | 0.10 | -0.14 | **La-La** | -0.24 | -0.22 | **Si-Si** | -0.06 | -0.06 |
| **Ca-Li** | 0.04 | 0.00 | **La-Mn** | -0.28 | 0.06 | **Si-Zn** | 0.00 | -0.04 |
| **Ca-Mn** | 0.07 | 0.04 | **La-Nd** | -0.12 | -0.21 | **Si-Zr** | -0.24 | 0.02 |
| **Ca-Nd** | 0.12 | -0.14 | **La-Sc** | 0.01 | -0.13 | **Sm-Sm** | 0.01 | -0.18 |
| **Ca-Sc** | 0.12 | -0.08 | **La-Sm** | -0.08 | -0.20 | **Sm-Y** | 0.05 | -0.17 |
| **Ca-Si** | -0.16 | 0.10 | **La-Y** | 0.00 | -0.19 | **Sm-Zn** | -0.08 | 0.07 |
| **Ca-Sm** | 0.13 | -0.13 | **La-Zn** | -0.09 | 0.09 | **Sm-Zr** | -0.05 | -0.16 |
| **Ca-Sr** | 0.10 | -0.12 | **La-Zr** | -0.10 | -0.17 | **Sn-Sn** | 0.09 | -0.03 |
| **Ca-Y** | 0.14 | -0.12 | **Li-Li** | 0.01 | 0.00 | **Sn-Y** | -0.14 | 0.05 |
| **Ca-Zn** | -0.06 | 0.06 | **Li-Mn** | 0.06 | -0.01 | **Sn-Zn** | 0.04 | -0.01 |
| **Ca-Zr** | 0.13 | -0.10 | **Li-Si** | -0.03 | 0.00 | **Sn-Zr** | -0.11 | 0.02 |
| **Dy-Dy** | 0.02 | -0.15 | **Li-Y** | 0.10 | -0.01 | **Sr-Sr** | 0.07 | -0.16 |
| **Dy-Er** | 0.02 | -0.14 | **Li-Zn** | -0.02 | -0.01 | **Th-Th** | -0.40 | -0.27 |
| **Dy-Gd** | 0.02 | -0.16 | **Li-Zr** | 0.15 | -0.03 | **Th-Zn** | -0.09 | 0.08 |
| **Dy-La** | -0.04 | -0.19 | **Mn-Mn** | 0.02 | 0.01 | **Th-Zr** | -0.28 | -0.17 |
| **Dy-Mn** | -0.21 | 0.03 | **Mn-Nd** | -0.11 | 0.04 | **Ti-Ti** | -0.36 | -0.10 |
| **Dy-Nd** | 0.01 | -0.17 | **Mn-Pb** | 0.10 | 0.01 | **Y-Y** | 0.06 | -0.16 |
| **Dy-Sc** | 0.01 | -0.11 | **Mn-Sc** | -0.29 | 0.00 | **Y-Zn** | -0.07 | 0.06 |
| **Dy-Sm** | 0.02 | -0.17 | **Mn-Si** | -0.35 | -0.07 | **Y-Zr** | -0.05 | -0.15 |
| **Dy-Y** | 0.04 | -0.16 | **Mn-Sm** | -0.14 | 0.04 | **Zn-Zn** | 0.00 | -0.03 |
| **Dy-Zn** | -0.08 | 0.06 | **Mn-Sn** | -0.01 | -0.01 | **Zn-Zr** | -0.05 | 0.02 |
| **Dy-Zr** | -0.08 | -0.15 | **Mn-Sr** | 0.13 | 0.06 | **Zr-Zr** | -0.30 | -0.14 |

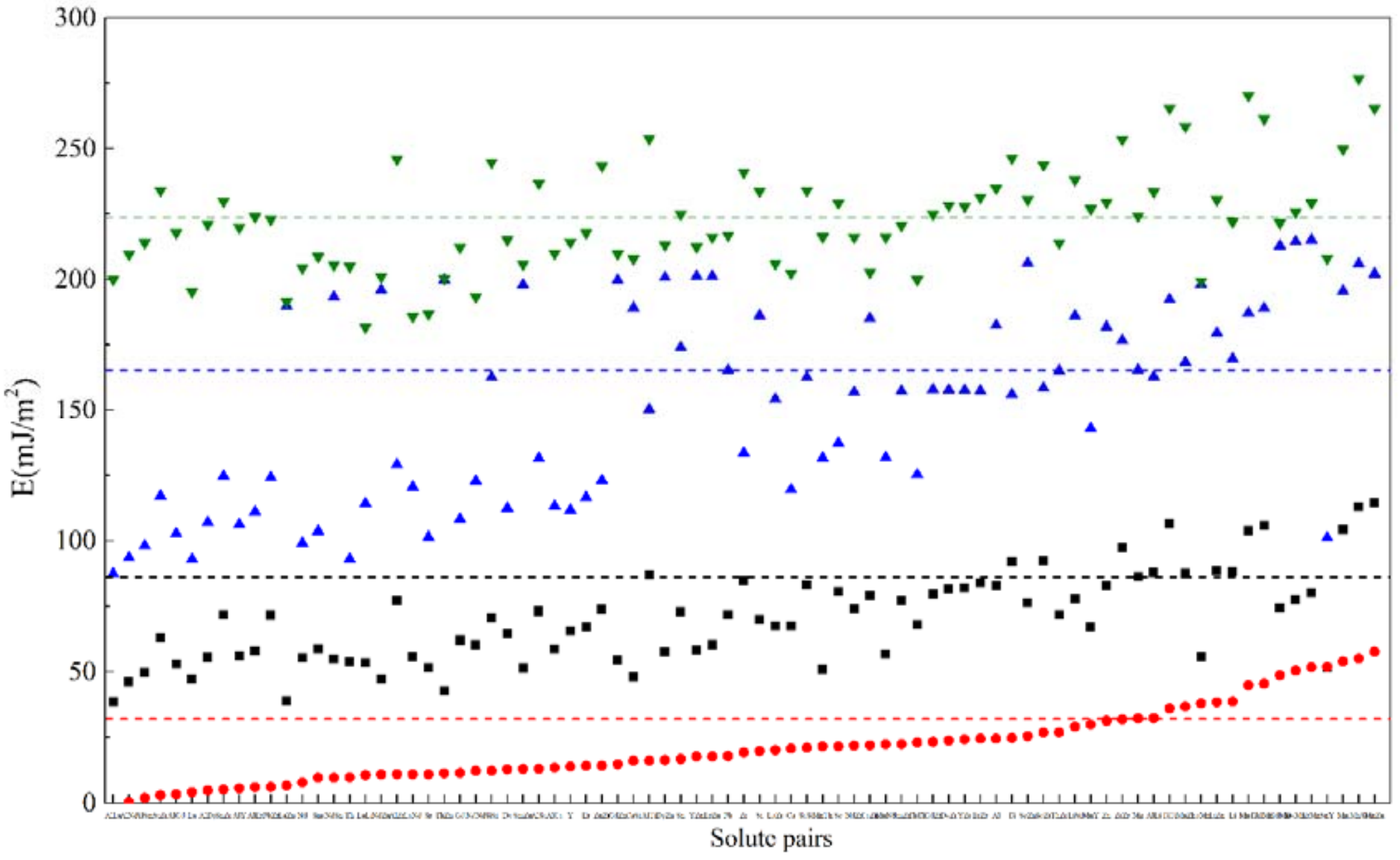

Figure 13 Basal $E_{sf}$ (red points), basal $E_{usf}$ (black points), prismatic $E_{usf}$ (blue points), and Pyramidal II $E_{usf}$ (green points) of Mg with possible solute pairs

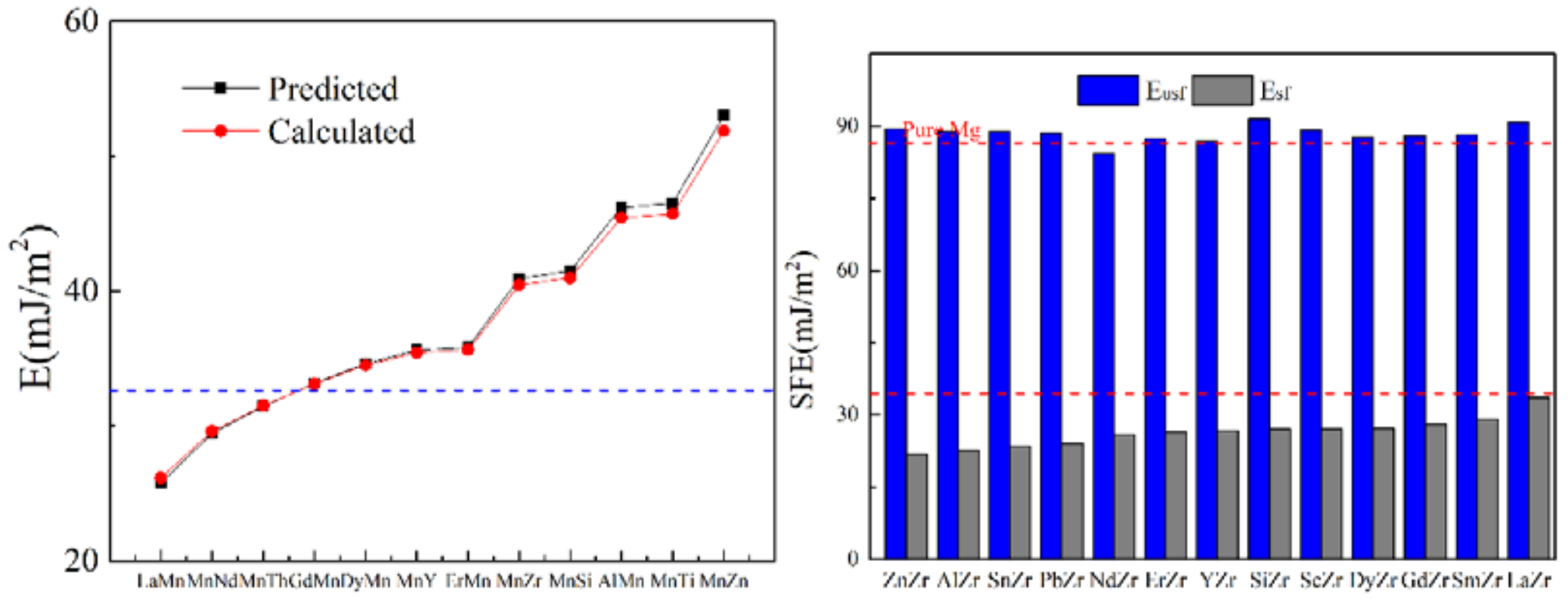

Figure 14 The calculated and predicted SFE of Mg-Mn-X and Mg-Zr-X series ternary alloys.

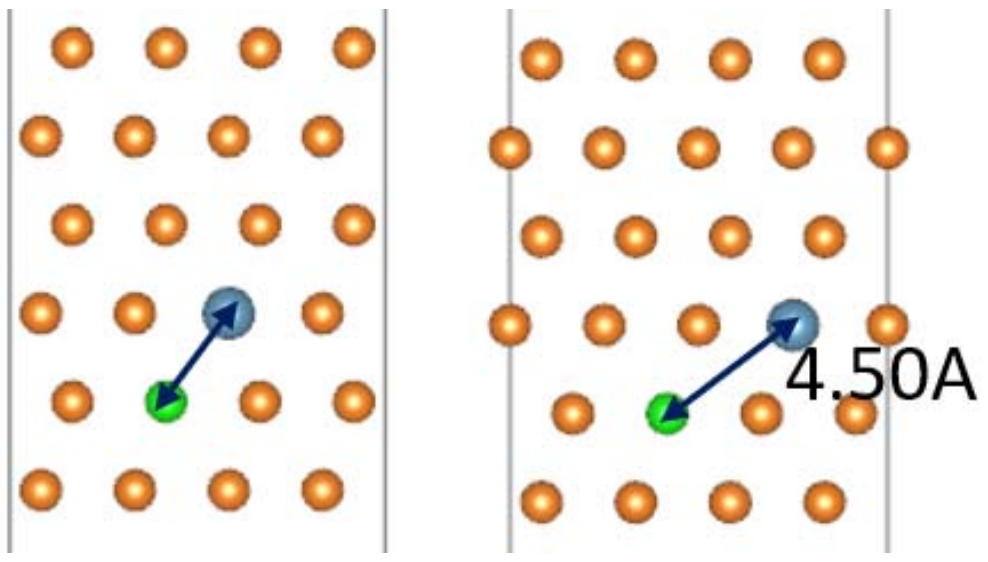

Figure 15 Configuration of initial state and stable SF state of basal slip under config δ.